\documentclass[jcp,preprint]{revtex4-1}
\usepackage{graphicx}
\usepackage{epstopdf}
\usepackage{amsmath}
\usepackage{hyperref}

\begin{document}
\draft
\draft
\title{The vibrational density of states of a disordered gel model}
\author{Lorenzo Rovigatti} 
\affiliation{ {Dipartimento di Fisica, Universit\`a di Roma {\em La Sapienza}, 
Piazzale A. Moro 2, 00185 Roma, Italy} }
\author{Walter Kob } 
\affiliation{Laboratoire Charles Coulomb,  
Universit\'e Montpellier 2, CNRS UMR 5221, 34095 Montpellier, France }
\author{Francesco Sciortino} 
\affiliation{ {Dipartimento di Fisica and  CNR-ISC, Universit\`a di Roma 
{\em La Sapienza}, Piazzale A. Moro 2, 00185 Roma, Italy} }

\begin{abstract}
We investigate the vibrational density of states (vDOS) in harmonic
approximation of a binary mixture of colloidal patchy particles with
two and three patches for different relative compositions $x_2$.
At low temperature, this system forms a thermo-reversible gel, i.e. a
fully bonded network of chains of two-patches particles, in which the
branching points are provided by three-patches particles.  For all
the compositions we find in the vDOS a pronounced peak at low frequency
whose height grows on increasing the fraction of two-functional particles, or
equivalently with the  average length of the chains.  To identify the
various spectral features,  we compare the vDOS of the whole system with
the one of small representative structures of the network and with the
vDOS of a long linear chain of two-patches particles and we find that
these structures are indeed able to rationalize the various peaks in
the vDOS of the full system.  At large $x_2$ the vDOS of the gel and of
the long chain show remarkable similarities. Analyzing the dispersion
relations and the spectrum of the linear chain we show that the
excess of low frequency modes, the analog of the boson peak in glassy
disordered systems,  arises from the strong coupling between  rotations
and translations.

\end{abstract}

\maketitle
\section{Introduction}
\label{section:intro}

The study of the vibrational density of states (vDOS) is central in
the investigation of solids~\cite{ashcroft_mermin_76,kuzmany_02},
since, apart from the structure, it is probably the most relevant and
direct quantity to characterize a system. E.g., within the harmonic
approximation, the vDOS (which we will denote by $g(\omega)$, where
$\omega$ is the frequency), allows one to calculate the temperature
dependence of the specific heat, thermal conductivity, as well as the
free energy. It has been known for a long time that the vDOS of {\it
disordered} systems shows anomalies, i.e. has features which are not
observed in crystalline systems. A well-known example is the vDOS of
fractal aggregates for which one finds that $g(\omega)$ shows at low
$\omega$ a power-law dependence with a fractal exponent, instead of
the typical Debye law $\omega^{d-1}$, where $d$ is the dimensionality
of space~\cite{alexander_orbach_82,rammal_toulouse_83,courtens_87}.
Deviations from the Debye law are also often observed in atomic
and molecular glasses and have been in the focus of interest of many
studies~\cite{buchenau_84,sokolov_93,taraskin_97,courtens_01,horbach_01,theenhaus_01,nakayama_02,courtens_03,wyart_05,schirmacher_06,ruffle_08,schmid_08,shintani_08,ilyin_09,binder_11}.
The most remarkable anomaly is an excess in the
number of modes, as compared to the Debye prediction, at
low frequencies (i.e.~for an atomic system in the THz range), a feature which is
commonly named ``boson peak'' and which is responsible for the
anomalous behavior in the specific heat or thermal conductivity of
glasses~\cite{courtens_01,courtens_03,nakayama_02,binder_11}. Although
some evidence has been provided for the existence of a
correlation between the fragility of a glass-forming system and
the strength of its boson peak, other results suggest that no such
correlation exists~\cite{sokolov_93,sokolov_network,novikov_05}. It
has been proposed that this correlation has a structural origin
\cite{sokolov_network,shintani_08}, since strong glass-formers have an
open network structure (due to the presence of directional/ionic bonds)
and tend to have a strong boson peak, while fragile liquids have a structure that is
similar to the one of the hard sphere system, i.e. van der Waals like
interactions, and usually a weak boson peak. Similar results have been
found in the Se$_x$As$_{1-x}$ system for which neutron scattering shows that the
number of low frequency modes is related to the average coordination number of
the atoms~\cite{kamitakahara_91}.

While much effort has been devoted to the investigation
of dense glasses, much less is know for the case of
thermoreversible gels, i.e.~disordered arrested states at low
densities~\cite{zaccarelli_07}. One of the reasons for this is that
the realistic modeling of thermoreversible gels has become possible only
recently~\cite{zaccarelli_05,delgado_05,bianci_06,blaak_07,saw_09,zaccarelli_06,russo_11}. Models
include patchy particles of low valence, particles interacting via
competing short and long range interactions, three-body potentials, dipolar particles,
and others~\cite{zaccarelli_07}. In all these models, the interaction
potential makes that the particles have only a small coordination
number which in turn allows to generate very open disordered structures,
composed by chains of bi-coordinated particles joined by a small number
of three- (or more) coordinated particles which act as junctions in the
network. In the last years the structural and relaxational properties
of these thermoreversible gel models have been extensively investigated
in the attempt to deepen our understanding of the mechanisms responsible
for the slowing down of the dynamics with decreasing temperature and to
clarify the differences between gels and glasses~\cite{zaccarelli_07}.

In the present work we extend this line of investigations and focus
on the vDOS of one of these models, a binary mixture of particles
with, respectively, two and three attractive patches. By changing the
relative composition of the mixture we can change in a controlled way the
structural properties of the system~\cite{russo_09} and, in particular,
the length of the bi-functional chains connecting the three-functional
particles.  This flexibility will allow us to study how a change of the
structure influences the properties of the vDOS as well as the behavior of
the system at low frequencies, where a strong coupling between rotational
and translational degrees of freedom  is found.

\section{Model and computational details}
\label{section:model}

The model used in the present study has been presented in detail in
Ref.~\cite{russo_09}. Hence we provide here only the necessary amount
of information required to make the article self-contained.

\subsection{The interaction potential}

The particles are modeled as spherical rigid bodies of diameter $\sigma$
and mass $m$. The surface of particle $i$ is decorated with $M_i$ sticky
spots called patches. These patches are located at a distance $\sigma /
2 $ from the particle's center of mass. The total interaction between
particles $i$ and $j$ is

\begin{equation}
V(i, j) = V_{CM}(i,j) + V_P(i,j),
\label{eq:potential}
\end{equation}

\noindent
where $V_{CM}$ is the interaction between the centers of mass of the
two particles, while $V_P$ is the potential acting between the patches.
So the total potential of the system is $V_{tot}=\sum_{i,j} V(i,j)$. We
have chosen

\begin{equation}
V_{CM}(i, j) = \left( \frac{\sigma}{r_{ij}} \right)^{200}
\label{eq:cm_potential}
\end{equation}

\noindent
and 

\begin{equation}
V_P(i, j) = - \epsilon \sum_{p=1}^{M_i} \sum_{q=1}^{M_j} \exp\left[ -\frac{1}{2} \left( 
\frac{r_{ij}^{(pq)}}{0.12 \sigma} \right)^{10} \right] \quad ,
\label{eq:patchy_potential}
\end{equation}

\noindent
where $r_{ij}$ is the distance between the centers of mass of particles
$i$ and $j$ and $r_{ij}^{(pq)}$ is the distance between patch $p$
on particle $i$ and patch $q$ on particle $j$. With this choice of
parameters, the resulting potential resembles a hard sphere potential
complemented by an attractive square well-like potential if patches
on different particles are close to each other. In addition, the very
short range of $V_{P}$ guarantees that the so-called ``single bond per
patch condition'' is fulfilled. This condition means that the maximum
number of possible bonds that particle $i$ can form is equal to $M_i$,
and it allows to make an analytical description of the structural and
dynamical properties of the model~\cite{bianci_06,bianci_07,bianci_08}. By
choosing $\epsilon = 1.001$, one finds that the depth of the $V(i, j)$
potential is $u_0 = 1.0$.

\subsection{Details of the simulations}

The studied system is a  binary mixtures of $N_2$ bi-functional
particles, $M_i=2$, and $N_3$ three functional particles, $M_j=3$, with
$N \equiv N_2 + N_3 = 1000$ at packing fraction $\phi= \frac{\pi}{6}
\frac{\sigma^3 N}{V}=0.1$, where $V$ is the total volume of the system.
Patches are arranged on the poles for particles with two patches
(bi-functional particles) and equally spaced on the equator for particles
with three patches (three functional particles). We have studied six
different systems having the following concentrations of bi-functional
particles $x_2 \equiv N_2 / N$: $0.2$, $0.5$, $0.66$, $0.75$, $0.85$,
and $0.9$. This gives us the possibility to investigate how the average
valence $\overline{M} \equiv 3-x_2$ influences the structure and the
vibrational properties of the system.  In the following, energy is
measured in units of $u_0$, distance in units of $\sigma$, and time in
units of $\sqrt{m \sigma^2/\epsilon}$.  Temperature, $T$, is measured
in units of energy, setting the Boltzmann constant $k_B = 1$.

The configurations we analyze were generated via Brownian
dynamics~\cite{russo_09} by equilibrating the systems at $T=0.055$
(a $T$ at which  essentially most of the possible bonds have already
been formed) and by quenching the final equilibrium configurations
to $T=0.04$.  The simulation at $T=0.04$ is performed until the few
remaining isolated monomers attach to the spanning cluster. Indeed, at
such a low temperature,  the probability of breaking a bond, proportional
to $e^{-1/k_BT}$, is of order $10^{-11}$ and hence no bond breaking occurs
during the simulation ($\sim 10^8$ time steps, where each time step
correspond to $\delta t=0.001$). We stress that, in the resulting structure,
all the particles belong to the same cluster and the connectivity of the
system is not altered by the dynamics, i.e. in the following we study
only the vibrational dynamics and not relaxation and thus the system can
be considered as a chemical gel. Note that this vibrational dynamics is
not harmonic since the system is able to overcome local barriers, but which are not
associated to bond breaking events.

\subsection{Method to evaluate the harmonic density of states}

We have computed the normal modes and eigenfrequencies
of the system by determining the properties of the local
potential energy minimum configuration $\mathbf{q}^*$, i.e.
a configuration for which $ \left. \boldsymbol\nabla_\mathbf{q}
V_{tot}(\mathbf{q})\right\vert_{\mathbf{q}^*} = \mathbf{0} $,
where $\mathbf{q}$ are the generalized coordinates (center of
mass positions and orientational angles). This local minimum was
obtained by taking a configuration of the system as a starting point
of a minimization procedure in the potential energy $V_{tot}$
and the so obtained local minimum is $\mathbf{q}^*$. (Note
that this minimum configuration is often also called ``inherent
structure'' (IS)~\cite{stillinger_84,sciortino_05}.) In practice we have used
a conjugate gradient algorithm~\cite{numericalrecipes} which is here a fast
procedure since the system needs only to relax the thermal fluctuations
without altering the connectivity of the system.

Within the harmonic approximation, the local minimum of $V_{tot}$
can be written as a bilinear form in $\mathbf{q}_i$ and $\mathbf{q}_j$

\begin{equation}
V_{tot}'(\mathbf{q}) = 
\frac{1}{2} \sum_{i=1}^G \sum_{j=1}^G  \mathbf{q}_i \hat{\mathbf{H}}_{ij}  \mathbf{q}_j
\end{equation}

\noindent
where $G$ is the number of degrees of freedom and  

\begin{equation}
\hat{\mathbf{H}}_{ij} \equiv \left. \frac{\partial^2 V_{tot}(\mathbf{q})}
{\partial q_i \partial q_j} \right\vert_{\mathbf{q}^*}.
\label{eq:hessian}
\end{equation}

The eigenvectors of the matrix $\hat{\mathbf{H}}_{ij}$ give the normal
modes of our system and the square root of the eigenvalues give the
eigenfrequencies. The normalized distribution of the latter is thus the
vDOS $g(\omega)$.

To compute the Hessian $\hat{\mathbf{H}}_{ij}$ at the IS we have used
a finite difference method. Since bifunctional particles have patches
located on the poles, only two angles are needed to identify the
direction of the axis that joins them. This lowers the total number
of degrees of freedom from $6N-3$ to $5N_2 + 6N_3-3$.  The subsequent
diagonalization was performed using the LAPACK package. To improve the
statistics we have averaged over $20-50$ configurations.  Although these
configurations are not independent (since they share the same network
topology), this averaging significantly smooths the vDOS. We associate
such improvement to sampling of the several secondary minima composing
the same metabasin~\cite{heuer_08}.

\section{Results}

\subsection{Structure of the investigated systems}

The networks studied in this article are composed  by chains of
bi-functional particles that are cross-linked by the three functional
particles, acting as hinges between the chains.  Since bonds do not
break during the time scale of the simulation, the system behaves as an
almost fully connected chemical gel~\cite{russo_09}. The possibility of
tuning the concentration of junctions  allows us to generate structure
that have different chain lengths. Specifically, the average chain
length, i.e.  the average number of bi-functional particles between two
three functional ones (a length that is related to the
average distance between branching points of the network) is given  by~\cite{russo_09}

\begin{equation}
\bar{l} = \frac{3 - x_2}{3 - 3x_2}.
\label{eq:lbar}
\end{equation}

For the smallest value of $x_2$ studied here we have thus
$\bar{l}\simeq1.2$, while for the largest we obtain $\bar{l}=7.0$.
The ability to predict theoretically $\bar{l}$ is made possible by
the fact that the structure is essentially generated from an equilibrium
configuration and hence can be quite accurately described  by the theory
of Flory and Stockmayer\cite{bianci_07,flory_54}.

\begin{figure}
\centering
\includegraphics[width=7cm]{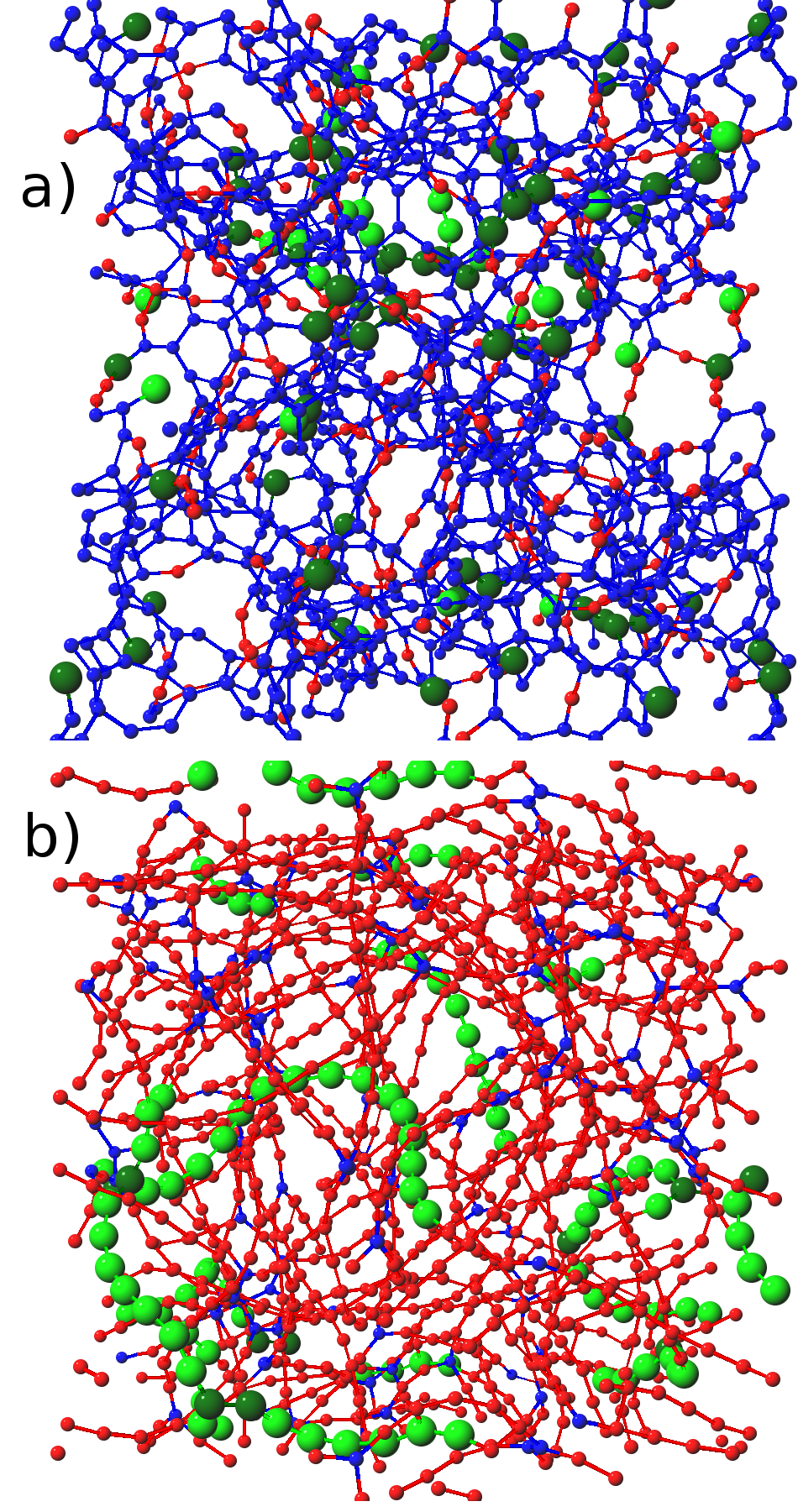}
\caption{A snapshot of the system at $T=0.04$ for $x_2=0.2$ (a) and
$x_2=0.9$ (b).  Small spheres: Bi-and three-functional particles with
low mobility (red and blue, respectively).  Large spheres: Bi-and
three-functional particles with high mobility (light and dark green,
respectively).  A particle is defined to be mobile if it has a mean
squared displacement $> 0.5$ for $x_2=0.2$ and $> 7.0$ for $x_2=0.9$.}
\label{fig1}
\end{figure}

To provide a visual representation of the networks generated for different
values of the mean connectivity $\overline{M}$, we show in Fig.~\ref{fig1}
two snapshots of the systems with $x_2=0.2$ and $x_2=0.9$, where the
most mobile particles are highlighted~\cite{footnote}. From the figure
we recognize that for $x_2$ small, i.e. high connectivity, the system
seems to be more heterogeneous than if $x_2$ is large. This is due to
the fact that in the former case the coexistence curve is closer to the
investigated state point and thus fluctuations are larger~\cite{russo_09}. 
Furthermore we see that for small average connectivity the
most mobile particles form chains whereas for small $x_2$ their spatial
distribution seems to be rather random. It is evident that the chain-like
structure found at low connectivity will give rise to low frequency
modes in the vDOS, and below we will discuss this in more detail.

\subsection{Density of states}

In this section we discuss the density of states, calculated within
the harmonic approximation, for different concentration of bi-functional
particles and correlate specific features of $g(\omega)$ with specific
vibrational modes of the system.

\begin{figure}
\centering
\includegraphics[type=pdf,ext=.pdf,read=.pdf,width=8cm]{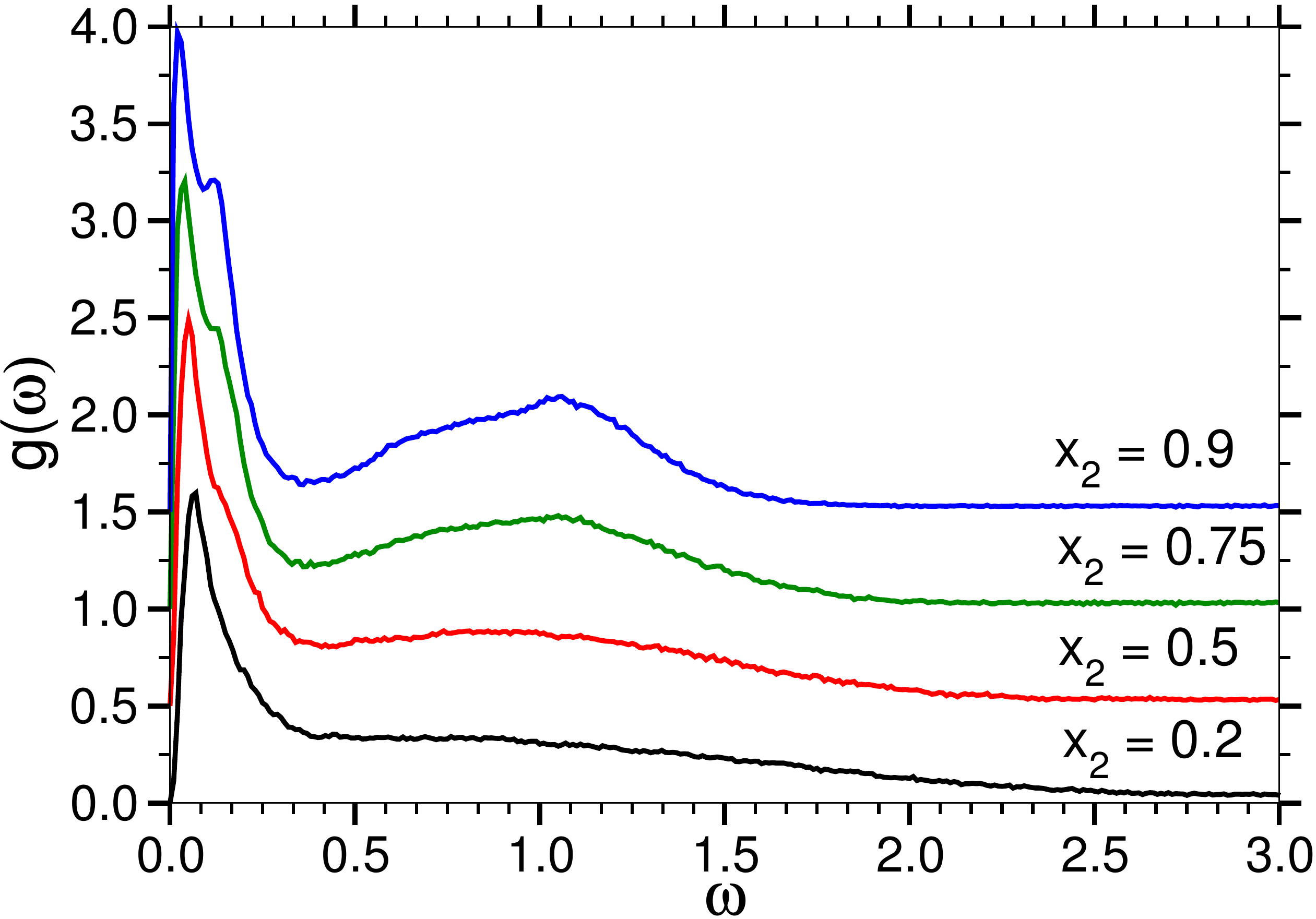}
\caption{Density of states $g(\omega)$ for different fractions of
bi-functional particles $x_2$. At the lowest $x_2$ there is one peak and
a long tail. With increasing $x_2$ a new peak appears while the minimum
between the peaks becomes more pronounced but remains at the same position
($\omega_{min} \simeq 0.38$). Consecutive curves are shifted vertically by 0.5.}
\label{fig2}
\end{figure}

The harmonic density of states for several values of $\overline{M}$
is shown in Fig.~\ref{fig2}.  For small $x_2$ the vDOS has one main
peak at small frequencies and a tail that extents to high $\omega$.
With intermediate and large $x_2$ the vDOS displays two peaks, one
located for $\omega \lesssim 0.38$ and one centered around $\omega
\approx 1$, followed by a broad tail extending up to $\omega \approx 10$.
Apart from the case $x_2=0.2$ (corresponding to $\bar l \approx 1.2$),
which shows a clear decrease in the number of low  frequency eigenmodes,
no significant dependence on $x_2$ is observed in the fraction of modes
contributing to the two main peaks and to the tail of the vDOS. This
suggests that increasing the length of the chains does not strongly
influence the total number of modes in the vDOS at low frequencies,
but only the shape of $g(\omega)$. Indeed, with increasing  $x_2$, the
low frequency peak  moves to lower frequencies and increases its height
while the peak at intermediate frequencies becomes better resolved  and a
shoulder  appears at $\omega \simeq 0.6$. For completeness, we note that a
shoulder is also present in the low frequency peak ($\omega \simeq 0.1$),
which turns into a minor peak when $x_2=0.9$. 

Note that the peak at low $\omega$ is observed at frequencies that are
significantly smaller than the {\it typical} frequencies of the model,
which are on the order of unity (due to the choice of the units). Since
the mentioned peak occurs at frequencies that are by a factor of 3-10
times smaller, we can tentatively identify this peak as a boson peak.
From the figure we thus can conclude that there is a clear correlation
between the {\it structure} of the system (open vs. compact network) and
the height of the peak (high/low).

To better understand the nature of the oscillations in the different
regions of the spectrum, we study the \textit{participation ratio}, $p_r$, 
of each  mode $i$. This quantity is defined as

\begin{equation}
p_r(\omega) = \left( N \sum_{n=1}^N \left\vert \mathbf{e}_i^n \right\vert^4 \right)^{-1},
\end{equation} 

\noindent
where $\mathbf{e}^n_i$ is the displacement vector of the $n$-th particle
of the $i$-th normal mode. The participation ratio is a standard measure
of localization of a mode\cite{laird_91,schober_96}. For
example, for a translation of the system $p_r(\omega) = 1$, while for
a vibration of a single particle $p_r(\omega) = 1/N$.

In dense systems the direction of the vibrations of a given particle,
i.e. $\mathbf{e}^n_i$, is basically independent of the direction of
the bonds that the particle forms with its neighbors. For a system that
has long chains, as the one studied here, this is no longer the case and
hence it is of interest to be able to discriminate between longitudinal
and transverse modes with respect to the bond direction (i.e. along the local
curvilinear direction of the chain). To this aim we quantify to what
extent  the displacement of the $i$-th normal mode is parallel to the
local direction of the bond.   By writing the  direction joining  bonded
particles $n$ and $j$ as $\mathbf{v}^n_j \equiv {\bf r}_{nj}/r_{nj}$, the
 observable $O(\omega)$  is thus  defined as

\begin{equation}
O(\omega) = \frac{1}{N} \sum_{n=1}^N \max_j \left\vert \mathbf{e}^n_i 
\cdot \mathbf{v}^n_j \right\vert^2,
\end{equation}

\noindent
where $j$ is an index running over the bonded neighbors of particle $n$. Since $\mathbf{v}^n_j$ as well as $\mathbf{e}^n_i$ are normalized,
$O(\omega) \in [0,1]$.  In the case of a unidimensional chain,
a longitudinal mode has $O(\omega) = 1$ while a transverse one has
$O(\omega) = 0$.

\begin{figure}
\centering
\includegraphics[type=pdf,ext=.pdf,read=.pdf,width=8cm]{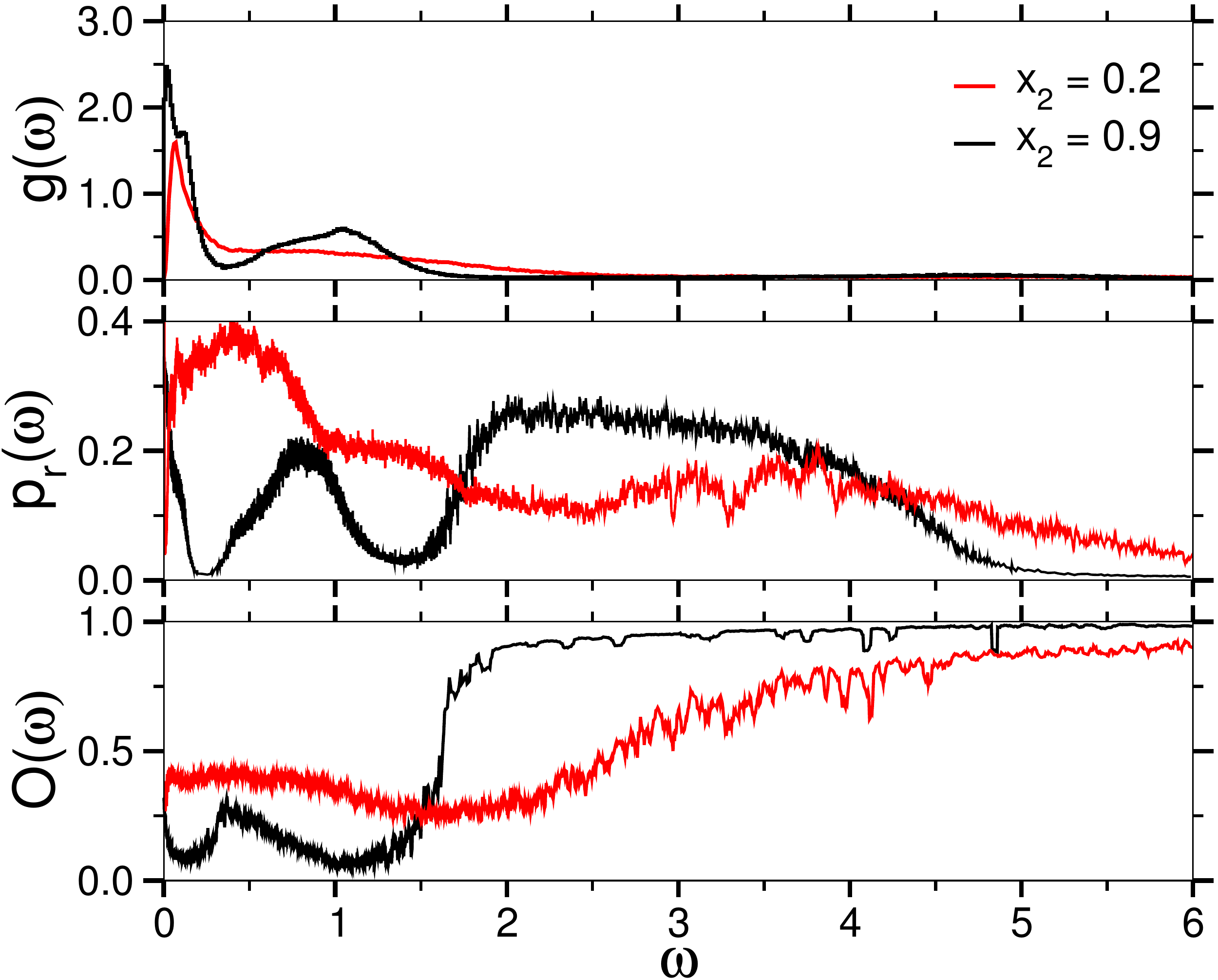}
\caption{$g(\omega)$ (top), $p_r(\omega)$ (middle), and $O(\omega)$ (bottom) for $x_2=0.2$ and $x_2=0.9$.}
\label{fig:eigen_pr_mode_bulks}
\end{figure}

The functions $p_r(\omega)$ and $O(\omega)$  are shown
in Fig.~\ref{fig:eigen_pr_mode_bulks} for $x_2=0.2$ and $x_2=0.9$, together with the
corresponding vDOS for an extended frequency range.  On increasing
$x_2$ both functions become more structured, suggesting that different
frequency regions are populated by modes with different characteristics.
We start by discussing the $x_2=0.9$ case. The data for $O(\omega)$
shows that modes for $\omega > 1.5$, i.e. modes that are not in the two
peaks seen in $g(\omega)$, have $O(\omega)\approx 1$, suggesting that
such modes are mostly vibrations along the bond directions (reminiscent
of longitudinal modes). In contrast to this, modes  with $\omega <
1.5$ are characterized by a rather small value of $O(\omega)$, thus
describing oscillations orthogonal to the chain direction, i.e. are
similar to transverse modes. Note that the transition of $O(\omega)
\approx 1$ to values $\leq 0.5$ at $\omega\approx 1.5$ is quite sharp,
indicating a well defined cross-over from longitudinal modes at high
frequencies to transverse modes at low $\omega$. The participation ratio,
which never exceeds the value $0.4$, suggests that all modes are mostly
localized. We recall that the value usually taken as indication of a
collective oscillation, i.e. of propagating phonons\cite{schober_96},
is $p_r \gtrsim 0.5$. Note that such propagating phonons are not even
seen at low frequencies where one does expect acoustic modes. The absence
of such excitations indicates that on the length scale of the simulation
box the system is still very heterogeneous and as a consequence the
modes cannot propagate. However, it can be expected that for larger
simulation boxes such propagating modes will indeed be present and the
rise of $p_r(\omega)$ at small $\omega$ can be taken as indication that
one is approaching this acoustic regime. Also at values of $\omega$ in
which $g(\omega)$ shows the peak around $\omega \approx 1.0$ we see that
$p_r(\omega)$ is significantly larger than zero. Thus we conclude that
also these modes show a significant delocalization. As we will show below,
these modes correspond in fact to excitations in which a chain makes
transverse oscillations, in agreement with the small values of $O(\omega)$ 
in this $\omega-$range. Last but not least we see that $p_r(\omega)$
is also relatively large if $\omega >2.0$. This is (probably) related to
the fact that at elevated frequencies the modes are a hybridization of
(relatively) localized modes, thus should not really be considered as
extended objects~\cite{schober_96}.

In the case of $x_2=0.2$, $O(\omega)$ never reaches very small values
and the difference between the region of the peaks and the tail is less
pronounced. This indicates that even modes with $\omega < 3 $ cannot be
described to be transverse (with respect to the bond direction).  This is
consistent with the very short average chain length observed for small
values of $x_2$. Furthermore we see that $p_r(\omega)$ becomes relatively
large as soon as $\omega \leq 1.0$, i.e. for these frequencies the modes
are significantly more collective than for the case $x_2=0.9$. This
is in agreement with the fact that for $x_2=0.2$ the structure of the
system involves smaller length scales than the ones for large values of $x_2$
and hence acoustic like modes can propagate even at relatively high
frequencies. From $O(\omega)$ we also recognize that at large frequencies the
modes appear to describe motion parallel to the bonds, i.e. $O(\omega)$ is
large. Since the participation ratio is relatively small, this suggests
that these longitudinal modes are localized.  This feature is thus
directly related to the structure of the system which, for small $x_2$,
shows short rings of bonded particles, as displayed in Fig.~\ref{fig1}a.

\begin{figure}
\centering
\includegraphics[type=pdf,ext=.pdf,read=.pdf,width=8cm]{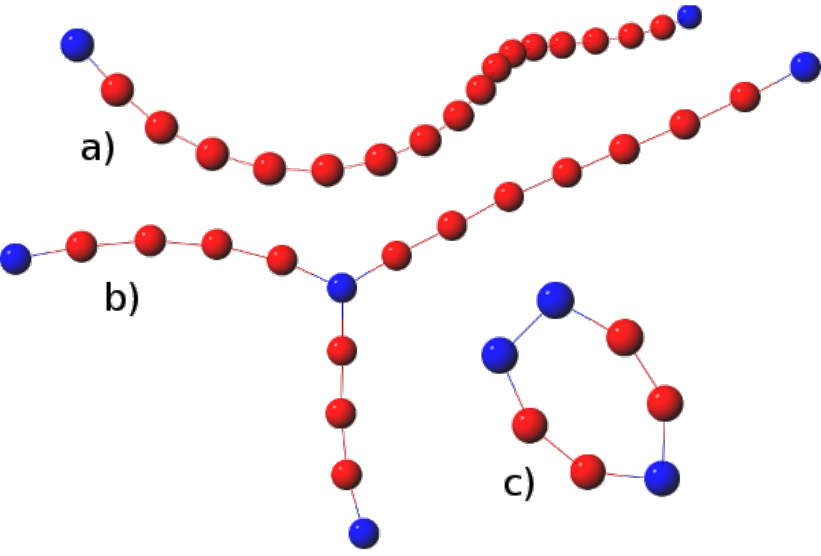}
\caption{Studied network elements: a) 
A curved chain formed by 18 particles; b) Three chains linked together, 18 
particles in total; c) A closed ring. All these structures have been found 
in $x_2=0.9$ configurations at $T=0.04$.}
\label{fig:structures}
\end{figure}

To gain insight into the nature of the modes in the different regions
of the spectrum we compare the vDOS of the system with the vDOS of
isolated elements of the network, found in the simulated configurations.
In particular we focus on  three different structures: A bended chain
of bi-functional particles, a star configuration in which three chains
of bi-functional particles are connected via a central three-functional
particle and a ring of bi- and three-functional particles. The first
two structures are typical elements for the network with large $x_2$,
while the ring is relevant of the case with small $x_2$. Figure
\ref{fig:structures} shows the three selected structures, composed
respectively of eighteen, eighteen and seven particles.  The large number
of particles in the first two structures allows us to explore a larger and
better resolved range of frequencies. For the chain the number of degrees
of freedom, and therefore the number of eigenvalues obtained, is $92$;
for the star structure it is $94$, and for the ring $38$. We have also
computed the density of states for chains whose length is comparable to
the average chain length ($\bar{l}\simeq 7$) and we have obtained the same
qualitative results~\cite{rovigatti_09}.

\begin{figure}
\centering
\includegraphics[type=pdf,ext=.pdf,read=.pdf,width=8cm]{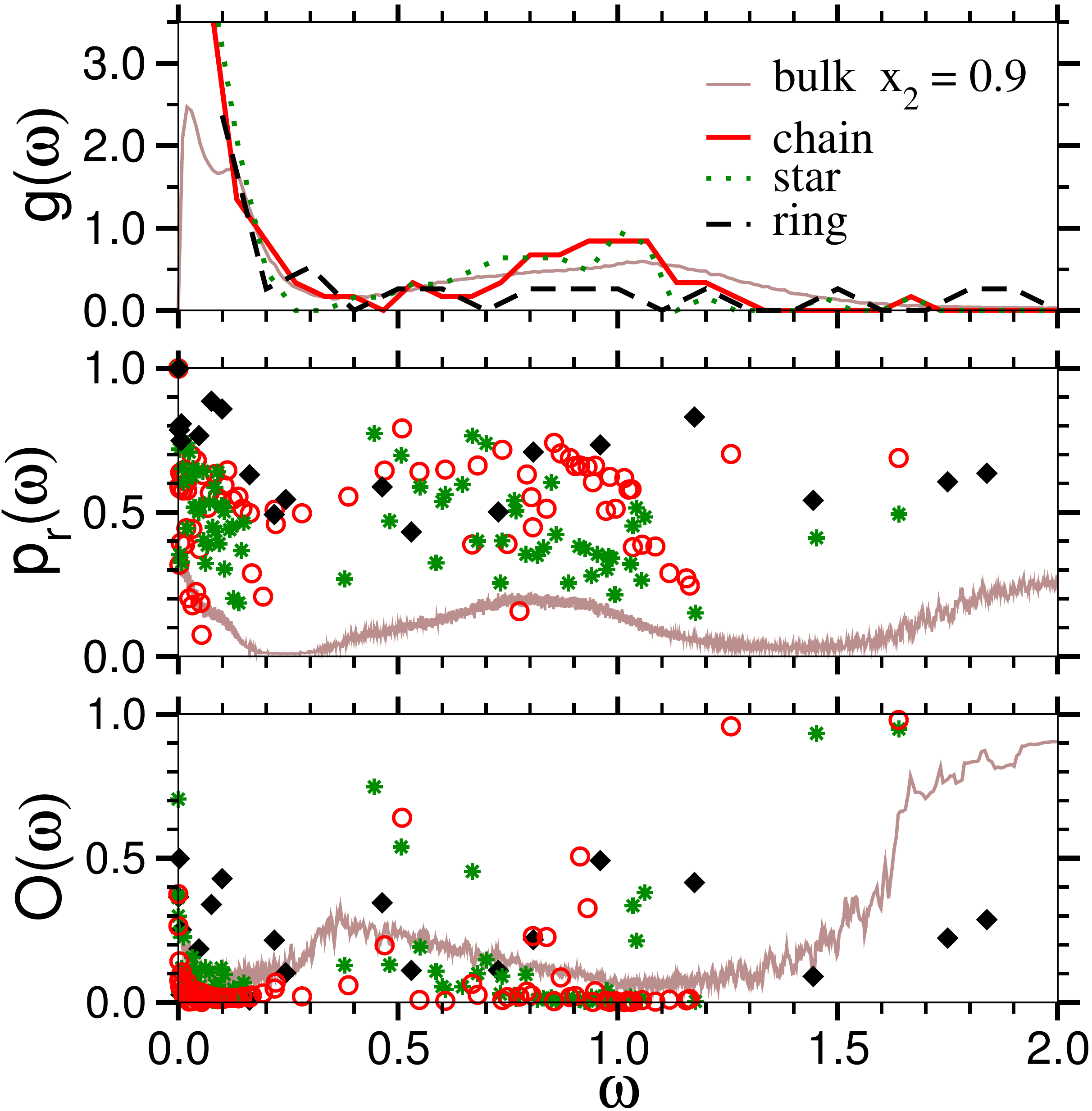}
\caption{$g(\omega)$ (top), $p_r(\omega)$ (middle), and $O(\omega)$
(bottom) for the structures pictured in Fig.~\ref{fig:structures}: a ring (filled diamonds), a 
star of chains (stars) and a chain (open circles). For
reference we include also the $g(\omega)$, $p_r(\omega)$ and $O(\omega)$ of the entire
system for $x_2=0.9$.}
\label{fig:structures_eigen_pr_mode}
\end{figure}

For each of these structures we have computed the density of
states after the usual minimization stage. The results are shown
in Fig.~\ref{fig:structures_eigen_pr_mode}. The chain and the star
configurations have density of states that are remarkably similar to
the one of the system with $x_2=0.9$.  Instead, the ring's $g(\omega)$,
is qualitatively more  similar to the density of states of the system
with $x_2=0.2$, i.e. it does not have a pronounced peak at $\omega\approx 1.0$. 
Figure~\ref{fig:structures_eigen_pr_mode} also shows the
participation ratio and the $O(\omega)$ for the selected structures.
Some of the features found in the bulk system are also observed in
the eigenmodes of the examined structures, in that for $\omega> 1.5 $,
$O(\omega) \approx 1$ for the chain and the star while $O(\omega) \approx
0.2$  for the ring. Moreover, in the frequency region $\omega < 0.25$,
both $O(\omega)$  as well as  $p_r(\omega)$ are larger for the ring than
for the chain and the star.  Again these features are very similar to the
ones found  when analyzing the $x_2=0.2$ and $x_2=0.9$ systems.  Hence,
despite the small number of degrees of freedom of these structures,
a strong correspondence with  the features of the vDOS, of $O(\omega)$,
and $p_r(\omega)$ of the bulk system  can be found.  Thus the chain and
star are indeed representative of the low valence network, while the
ring is characteristic of the network rich in three-functional junctions.

\begin{figure}
\centering
\includegraphics[width=8cm]{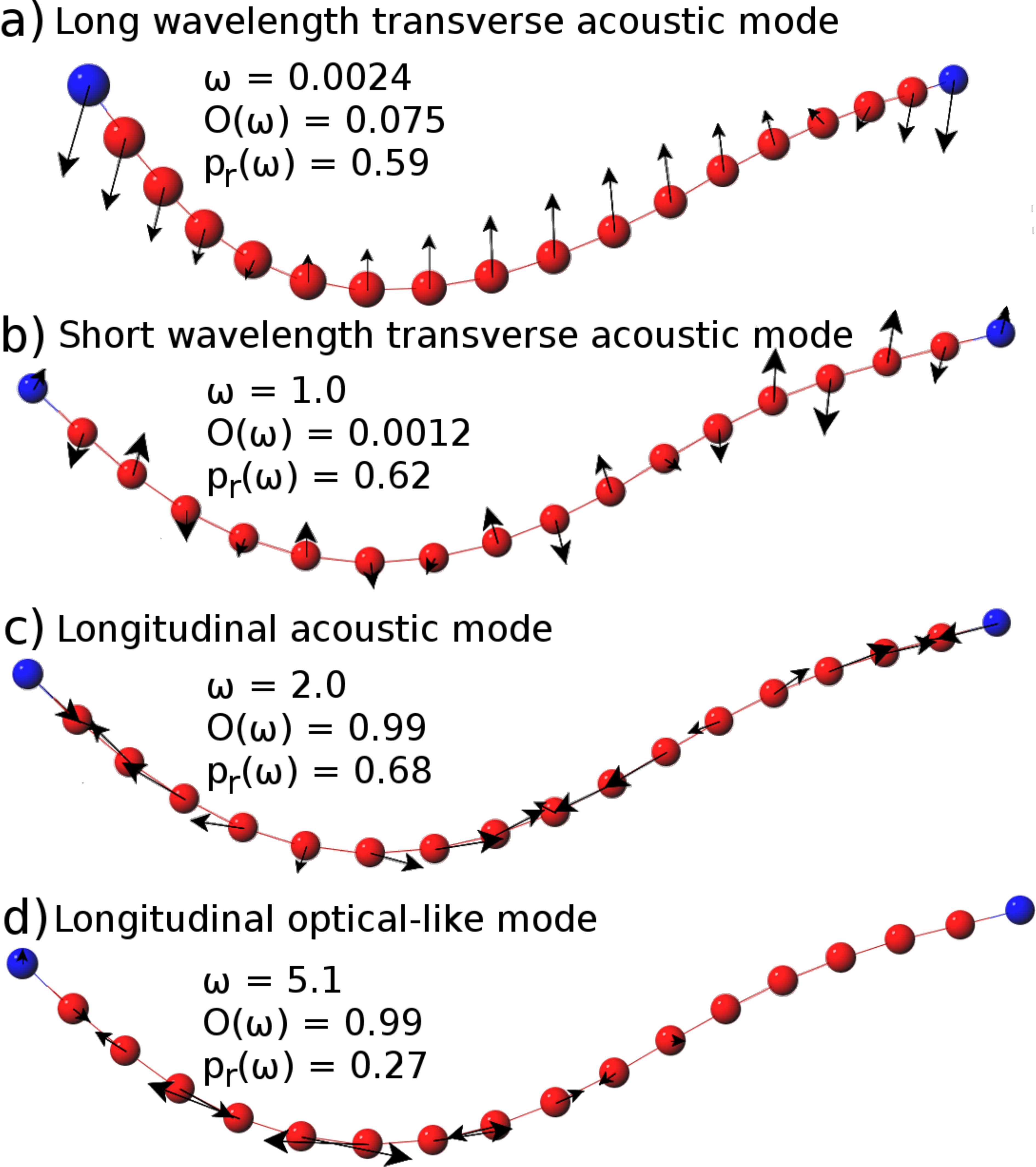}
\caption{Four characteristic vibrational modes of the bended chain.}
\label{fig:chain_modes}
\end{figure}

As a consequence of the observed similarity between the chain and the
$x_2=0.9$ system we have the possibility to identify the different modes
appearing in the vDOS. Figure~\ref{fig:chain_modes} shows graphically the
direction of the translational displacement of all particles belonging to
the chain for characteristic frequencies that are located in different
regions of the vDOS.  One can identify the progression typical of one
dimensional chains on increasing frequency, even if the purity of the
modes is significantly altered by the curvilinear structure.  Indeed,
the graphical inspection of the displacements, as well as the specific
values of $p_r(\omega)$ and $O(\omega)$, consistently suggest a strong
resemblance with a long wavelength acoustic transverse phonon  (a),  a
short wavelength acoustic transverse phonon  (b), a longitudinal acoustic
phonon (c) and an optical-like longitudinal phonon (d).  Of course,
the curvilinear nature of the chain introduces some disorder that alter
the pureness of the modes.

\begin{figure}
\centering
\includegraphics[width=8cm]{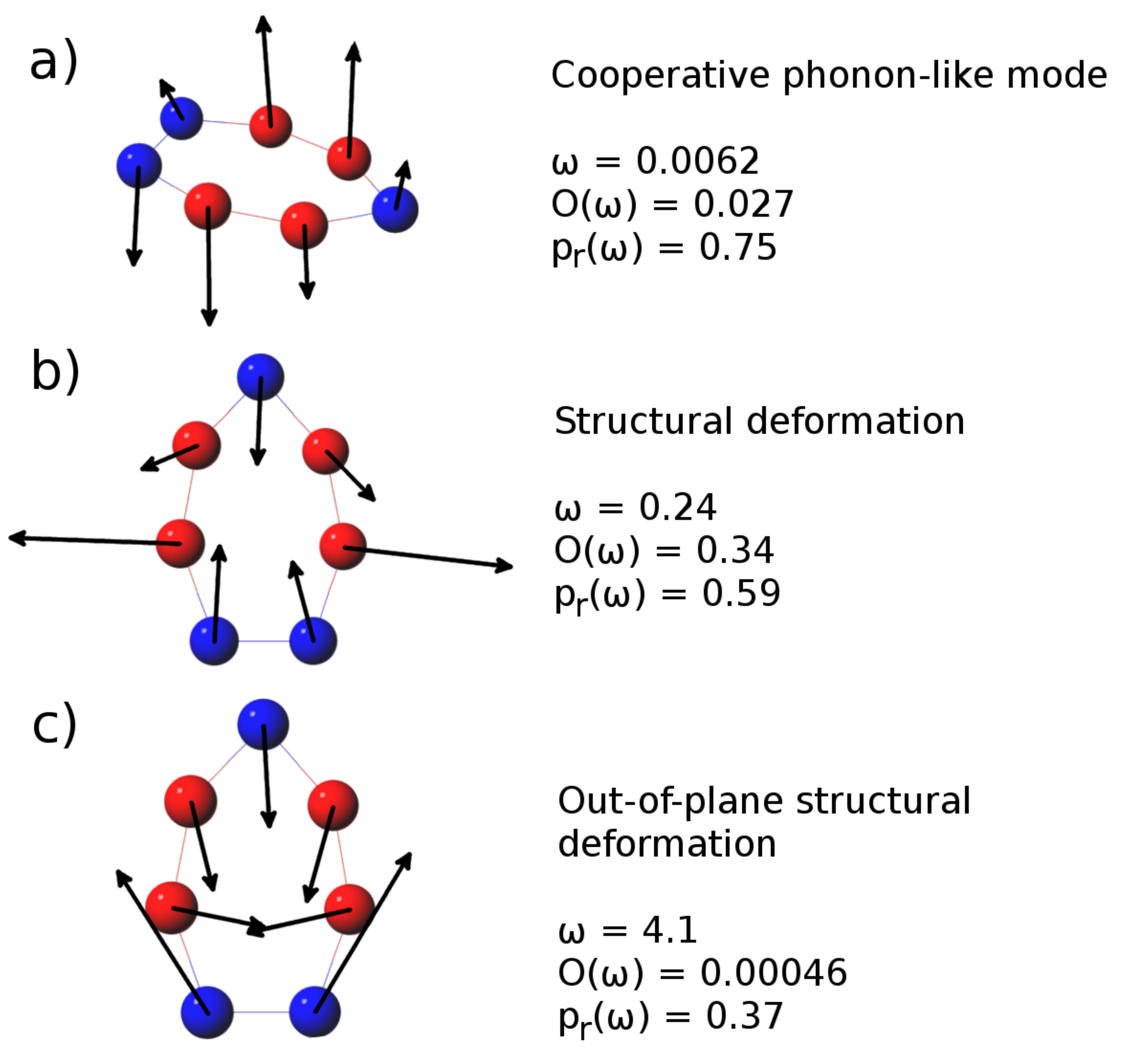}
\caption{Three characteristic vibrational modes of the ring.}
\label{fig:epta_modes}
\end{figure}

Finally, we show in Fig.~\ref{fig:epta_modes} three different modes
of the ring, located in different frequency regions.  Apart from the
low frequency mode which has a large participation ratio, the other
modes describe deformations of the structure. The different values of
$O(\omega)$ indicate if the deformation takes place along the plane
of the ring or if it is mostly perpendicular to it.

\begin{figure}
\centering
\includegraphics[type=pdf,ext=.pdf,read=.pdf,width=8cm]{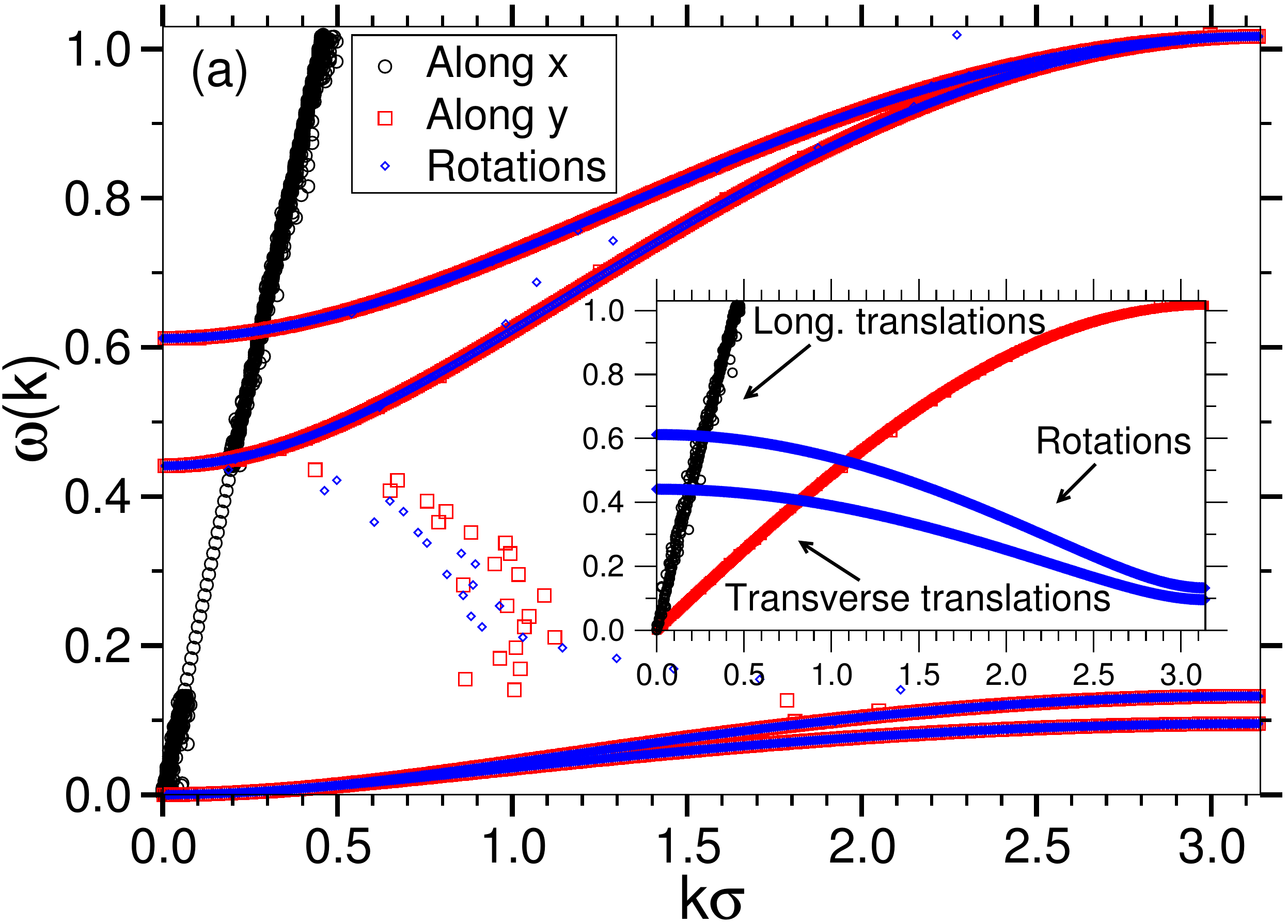}
\includegraphics[type=pdf,ext=.pdf,read=.pdf,width=8cm]{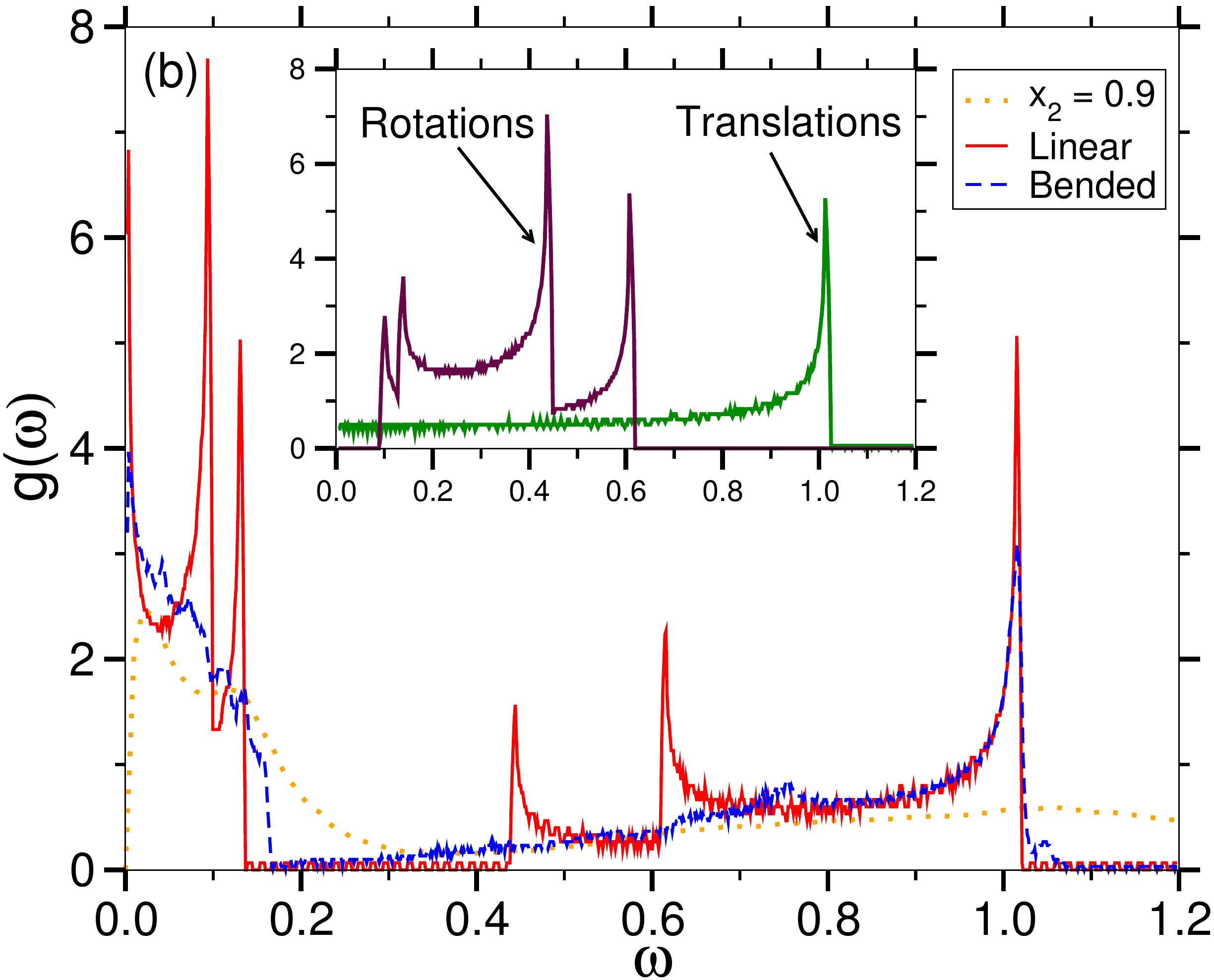}
\caption{
(a) Dispersion relation for the linear chain, calculated by estimating
the spacial periodicity along the $x$ direction (longitudinal) and along
the $y$ or $z$ (transverse) direction. The dispersion relations obtained by evaluating
the periodicity for the two angles coincides with the the $y$ or $z$
results, confirming the mixed nature of these modes.  The inset shows
the corresponding dispersion relation curves for the decoupled system
in which we diagonalize  only the $3N \times 3N$ block of the Hessian
associated to translational degrees of freedom as well as the  $2N \times
2N$ block associated to rotational degrees of freedom.
(b) vDOS of a linear and of a bended chain of $N=1000$ particles. The
vDOS of the network with $x_2=0.9$ is also shown for comparison. The
inset shows the vDOS calculated for the decoupled system.
Note that the figure shows only the region $\omega<1.2$
and therefore the van Hove singularity associated with the longitudinal
modes and which is located at $\omega \approx 4.4$ is not visible.
}
\label{fig:vDosxt}
\end{figure}

In order to further understand the origin of the peaks in the vDOS of our
gel, it is useful to compare it to the one of a one-dimensional linear
chain in the limit of very large $N$.  For this we have calculated the
vDOS for a system of 1000 bi-functional particles located on a straight
linear chain oriented in the $x-$direction. Neighbouring particles $i$
and $j$ are located at the distance corresponding to the minimum of
$V(i,j)$,  where  $V(i,j)$ is the same inter-particle potential used in
the simulation of the gel, see Eq.~(\ref{eq:potential}). The presence
of  angular degrees of freedom, which strongly couple to translational
degrees of freedom, significantly alters the behavior of the vDOS with
respect to the well-known text-book case of the one dimensional chain of
particles interacting only via harmonic springs~\cite{ashcroft_mermin_76}.
To evaluate the role of the coupling, we first start by studying the
corresponding system in which the coupling is missing. This is realized
by setting for the diagonalization the translation-orientation blocks
of the Hessian to zero~\cite{sciortino_94}.  Beside the translational
and rotational vDOS, we have also investigated the dispersion relation,
by analyzing the periodicity of the different eigenvectors. For this we
have calculated the space Fourier transform of each eigenvector and have
associated the position of the peak in $k-$space to the corresponding
eigenvalue $\omega$. In the case of translation, motion along the
chain direction provides information on the longitudinal branch, while
displacements in the two orthogonal directions provide information on
the transverse branches.  Similarly, the periodicity of the two angular
displacements of the particles along the chain provide information on the
rotational modes. The dispersion relations and vDOS for the decoupled system
are shown in the insets of Fig.~\ref{fig:vDosxt}a and b.  The longitudinal
and transverse sound velocities $v_l$ and $v_t$, calculated from the
slopes of the dispersion curves at small $k$ (or equivalently from the
square root of the second derivative of the potential with respect to
the chain direction $x$ for the longitudinal modes and with respect to
$y$ or $z$ for the transverse ones) are found to be $v_l\simeq2.24$ and
$v_t\simeq0.51$.  The significantly smaller value of $\frac{\partial^2
V(i,j) }{\partial y_i \partial y_j}$  as compared to $\frac{\partial^2
V(i,j) }{\partial x_i \partial x_j }$ gives rise to a $v_t$ that is
much smaller than $v_l$ and hence a transverse dispersion relation
located at frequencies much smaller than the longitudinal one. This
strong difference is plausible since it is much easier to deform
the chain in its orthogonal direction than in its parallel direction.
The translational vDOS of the system, inset of Fig.~\ref{fig:vDosxt}b, 
shows a flat density of states,
followed by a van Hove singularity around $\omega \approx 1$ (the
boundary zone of the transverse band) and around $\omega \approx 4.5$
(the boundary of the longitudinal band, not shown in the figure). 
The rotational modes have an optical character in that they have, at $k=0$, 
a finite frequency (inset of Fig.\ref{fig:vDosxt}a). With 
increasing wave-vector the modes become softer,
before they degenerate at the zone boundary.

Next we come back to the original model, i.e. we include the coupling
between translations and rotations when the Hessian is diagonalized.
We see that this coupling induces strong changes in the dispersion
relations, see main panel of Fig.~\ref{fig:vDosxt}a. The coupling makes
that there are now two transverse optical modes that at $k=0$ have the
frequency of the rotational modes of the uncoupled system. Similarly the
coupling makes that the rotational degrees of freedom can also have an
acoustic branch, with a frequency at the zone boundary that is given by
the $\omega$ of the uncoupled system. In contrast to the rotational and
transverse modes, no significant change in the longitudinal excitations
are seen. As a result of the coupling between the various modes, the vDOS
at low $\omega$ becomes populated by modes which arise from the strong
coupling between the low frequency transverse translational modes and
the optical rotational modes (Fig.~\ref{fig:vDosxt}b), thus explaining
the presence of high intensity at low frequencies.

We have also calculated the vDOS of a long bended chain, and the
results are also reported in Fig.~~\ref{fig:vDosxt}b. We see that the features
of the linear chain are retained in the vDOS of the bended chain,
even if the  van-Hove singularities are now smeared out by the absence
of linearity. Interestingly enough, the main features of the linear chain
are also retained in the bulk system, whose broad bands can qualitatively
associated to  the linear chain bands, providing a strong  support for
the interpretation of the  disordered network modes  discussed previously
(see Fig.~\ref{fig:eigen_pr_mode_bulks}).

\section{Conclusions}
In this article we have studied in detail the vDOS of a disordered
system composed by poly-disperse chains of connected bi-functional
particles, joined together via three-functional junctions.  The relative
concentration of bi- and three- functional particles makes it possible
to control the average distance $\bar l $ between junctions and explore
the effect of the network mesh on the vDOS.

We have found a clear evolution of the vDOS with $\bar l$, with a
progressive structuring of $g(\omega)$.  To provide a characterization
of the modes in different frequency regions, we have calculated the
participation ratio and the mode amplitude  in  the bond direction,
which has allowed us to distinguish between localized and delocalized 
modes and transverse and longitudinal
excitations. We have found support for the resulting
classification via the analysis of specific structures of the networks,
i.e. chains, stars of chains and rings, finding that most of the features
of the bulk vDOS can be associated with specific modes of these simpler
structures.

We have also found very helpful the comparison with the vDOS of an
infinite chain of particles interacting with the same potential. The
comparison between the vDOS calculated in the presence and in the
absence of rotational-translational coupling helps clarifying the
importance of such a coupling and the nature of the bands observed
in the bulk disordered network.  In particular, we observed that the
coupling between translation and rotation significantly increases the
vDOS at low $\omega$, which becomes populated by transverse modes, with
frequencies which are strongly decreased  (as compared to the uncoupled
system) by the possibility of rotating the particles in coherence with
the translational displacements.  Interestingly, this band is clearly
retained also in the vDOS of the bulk system.

This study suggests that properties of the disordered systems can
partially be traced back to properties of the corresponding crystal, with
an adequate  smearing introduced by the absence of long range periodicity
and by the finite length of the chains. It also calls attention on the
important role of the coupling between translation and rotation in the
low frequency region of the spectrum, i.e. in the $\omega$ region where
the boson peak is observed in glasses. {\it For the present model} it
is this coupling that leads to a strong intensity of the vDOS at low
frequencies. However, this mechanism to generate low-frequency modes
is certainly not the only one, since, i.e., disorder in the coupling
constants or in the masses will also give rise to soft modes as has been
shown for other systems~\cite{schmid_08,ilyin_09}.

\section{Acknowledgements}

We thank J. Russo for help during the initial stages of this work and A.V. Dobrynin and U. Buchenau for helpful discussions.  LR and
FS acknowledge support from  ERC-226207-PATCHYCOLLOIDS and ITN-234810-COMPLOIDS. WK is a  senior member
of the Institut universitaire de France.


\begin{thebibliography}{99}
\bibitem{footnote}
In order to define the mobility of a particle we have calculated the
mean squared displacement for each particle and generated the
corresponding histogram. Note that this mean squared displacement is
bounded, since bonds do not break. We found that this distribution
has a very sharp peak at small displacements and a long tail at large
displacements~\cite{rovigatti_09}.
\end{thebibliography}
\end{document}